\documentclass[%
aip,
amsmath,amssymb,
reprint,nofootinbib%
]{revtex4-2}
\usepackage{graphicx,array,booktabs,rotating}
\usepackage{dcolumn}
\usepackage{bm}
\usepackage{mathptmx}
\usepackage{multirow}
\usepackage{afterpage,float,color,xcolor}
\usepackage{hyperref}
\usepackage{etoolbox} 
\usepackage{tabularx,tabulary}
\usepackage[capitalize]{cleveref}
\usepackage{times}
\hypersetup{
	colorlinks,
	linkcolor={blue!100!black},
	citecolor={blue!100!black},
	urlcolor={blue!100!black}
}
\newcommand{\PRLsep}{\noindent\makebox[\linewidth]{\resizebox{0.3333\linewidth}{1pt}{$\bullet$}}\bigskip}
\makeatletter
\newcommand\footnoteref[1]{\protected@xdef\@thefnmark{\ref{#1}}\@footnotemark}
\makeatother

\newcommand{\sk}[1] {\textcolor{black}{#1}}
\newcommand{\etal}{\textit{et al.}}
\usepackage[page]{totalcount}
\usepackage{etoolbox,fancyhdr,xcolor}%
\usepackage[displaymath,pagewise]{lineno}

\begin{document}
\preprint{AIP/123-QED}	
	\title[\textbf{Phys. Fluids} (2021) $|$  Manuscript Accepted \href{https://doi.org/10.1063/5.0075583}{[doi: 10.1063/5.0075583]}]{Unsteady pulsating flowfield over spiked axisymmetric forebody at hypersonic flows}

	\author{Mohammed Ibrahim Sugarno}
	\email{ibrahim@iitk.ac.in (Corresponding Author)}
	\affiliation{Department of Aerospace Engineering, Indian Institute of Technology, Kanpur 208016, India}%
	
	\author{R. Sriram}
	\affiliation{Department of Aerospace Engineering, Indian Institute of Technology - Madras, Chennai 600036, India}%
	
	\author{S. K. Karthick}
	\affiliation{Faculty of Aerospace Engineering, Technion-Israel Institute of Technology, Haifa-3200003, Israel}%
	
	\author{Gopalan Jagadeesh}
	\affiliation{Department of Aerospace Engineering, Indian Institute of Science, Bengaluru 560012, India}%
	
	\date{\today}
	
\begin{abstract}
\sk{The paper gives experimental observations on the hypersonic flow past an axisymmetric flat-face cylinder with a protruding sharp-tip spike. Unsteady pressure measurements and high-speed schlieren images are performed in tandem on a hypersonic Ludwieg tunnel at a freestream Mach number of $M_\infty = 8.16$ at two different freestream Reynolds numbers based on the base body diameter ($Re_D = 0.76 \times 10^{6}$, and $3.05 \times 10^{6}$). The obtained high-speed images are subjected further to modal analysis to understand the flow dynamics parallel with the unsteady pressure measurements. The protruding spike of length to base body diameter ratio of $[l/D]=1$ creates a familiar form of an unsteady flowfield called `pulsation.' Pressure loading and fluctuation intensity at two different $Re_D$ cases are calculated. A maximum drop of 98.24\% in the pressure loading and fluctuation intensity is observed between the high and low $Re_D$ cases. Due to the low-density field at low $Re_D$ case, almost all image analyses are done with the high $Re_D$ case. Based on the analysis, a difference in the pulsation characteristics are noticed, which arise from two vortical zones, each from a system of two `$\lambda$' shocks formed during the `collapse' phase ahead of the base body. The interaction of shedding vortices from the $\lambda$-shocks' triple-points, along with the rotating stationary waves, contributes to the asymmetric high-pressure loading and the observation of shock pulsation on the flat-face cylinder. The vortical interactions form the second dominant spatial mode with a temporal mode carry a dimensionless frequency ($f_2D/u_\infty \approx 0.34$) almost twice as that of the fundamental frequency ($f_1D/u_\infty \approx 0.17$). The observed frequencies are invariant irrespective of the $Re_D$ cases. However, for the high-frequency range, the spectral pressure decay is observed to follow an inverse and -7/3 law for the low and high $Re_D$ cases, respectively.}
\end{abstract}

\keywords{shock interactions, unsteady flow, hypersonic flow, modal analysis}

\maketitle

\section{Introduction}\label{sec:introduction}
Shock-shock\citep{Peng2020,Seshadri2020,Wang2018,Durna2016} and shock-boundary-layer\citep{Huang2020,Dlery2009,Ligrani2020} interaction is both complex and an interesting flowfield to study in the domain of supersonic/hypersonic aerodynamic research. In the past, several kinds of research \citep{Grasso2003,Mason2016,Gaitonde2015,Dolling2001} have been done to understand the nature of the flowfield, interacting shocks influences, and methods to control them\cite{Narayana2020,Desai2020,Deep2018,Wang2018}. An important aspect of such a flowfield is the unsteadiness induced by the interaction\cite{chen2019,Jain2021,Liu2021,Tekure2021,Zhou2021} which can either be local or global as in the case of self-sustained oscillatory flows \cite{Rockwell1979,Karthick2021,kumar2021,Xie2021}. The effects of local unsteadiness on the flowfield are minimal and can be mitigated using control techniques. However, the global unsteadiness specifically induced by the geometrical configuration significantly modifies the overall flowfield \cite{Mair1952,kabelitz1971}. One form of global unsteadiness is the flowfield over spiked forebodies\cite{bogdonoff1959,Maull1960584}.

\sk{Spiked forebodies were found to be very effective in reducing the aerodynamic drag in high-speed vehicles \citep{Mair1952,Maull1960584} among the other techniques. Rockets and missiles traveling at supersonic/hypersonic speeds within the atmosphere are subjected to severe aerodynamic heating and drag forces \citep{Ahmed2011,Karimi2019}. The problem of heating is minimized by having a blunt forebody; however, it comes with a penalty of increased drag \citep{Delery1999,Obrien1999}. Such blunt forebody shapes are only preferred for atmospheric re-entry vehicles where minimizing aerodynamic heating is a priority \citep{Braun1992}. However, in the case of vehicles traveling at high speeds within the atmosphere, reducing aerodynamic drag is of utmost importance to enhance its range and efficiency. The spiked forebody in any form, either in missiles\cite{Venukumar2006,Kulkarni2008} or engine intakes\cite{Sekar2020,Devaraj2020} reduces the overall drag as the spike creates a low pressure, recirculating, dead air region in front of the forebody. The pressure forces acting on the forebody are relatively lower than those acting on a forebody with no spike, resulting in drag reduction.} 

\begin{figure}
	\centering \includegraphics[width=0.8\columnwidth]{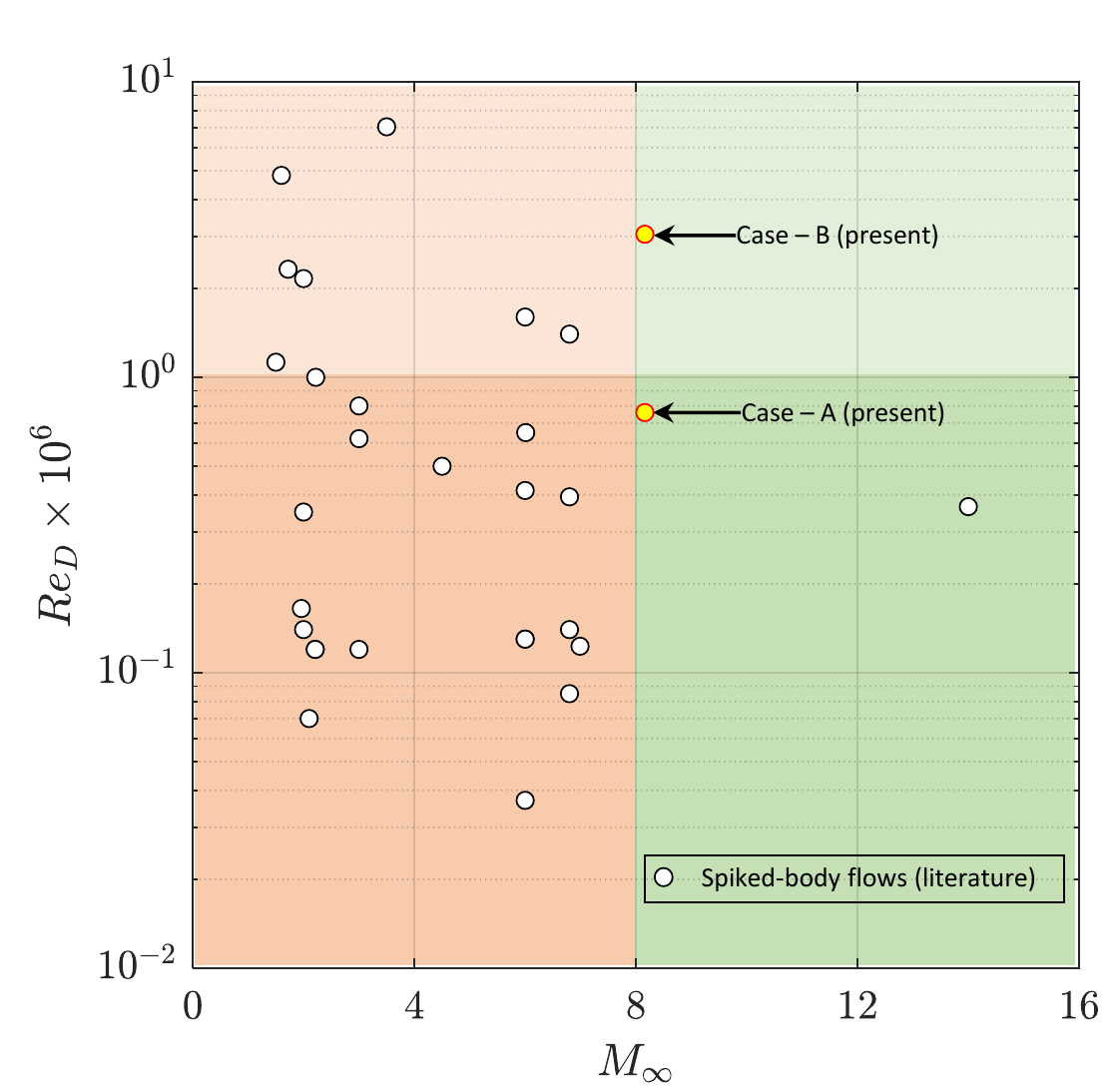}
	\caption{\sk{Graphical representation showing the list of spiked-body flow experiments performed over the last millennia as reported in the work of Sahoo \etal \cite{Sahoo2021} in the scatter plot format. The solid white-colored circles (black-outlined) show the freestream Mach number ($M_\infty$) versus freestream Reynolds number based on the base body diameter ($Re_D$) variations. The present experimental cases are marked as solid yellow colored circles (red-outlined). Orange and green shades demarcate the low and high $M_\infty$ regime, whereas the light and dark shades represent the corresponding zones of the low and high $Re_D$ regime.}}
	\label{fig:literature_cases}
\end{figure}

\sk{However, in spiked forebodies for certain range of $[l/D]$ ratios\cite{wood1962,holden1966,Kenworthy1978} (ratio of spike's length to base body diameter), the flowfield is either pulsating\cite{Feszty2004a} ($l/D < 1.4$) or oscillating\cite{feszty2004b} ($1.4 <[l/D]<2.5$). Pulsation is characterized by sudden collapse and rapid expansion of the forebody shock and separation region. In oscillation, the forebody shock changes its shape from concave to convex. Both these flow modes are highly unsteady, cyclic in nature, and geometry dependent, which has motivated many researchers to study them from the early 1950’s\citep{Panaras200969,Mair1952,Maull1960584,Antonov1976746}. A detailed list of studies on spiked body flows for a wide range of freestream Reynolds numbers based on the base body diameter ($Re_D$) and freestream Mach numbers are shown in Figure \ref{fig:literature_cases}. Most of them are experiments, and a few of them are computations. The plot is generated from the collective literature available on the spiked body flow physics presented in Table-1 of Sahoo \etal\cite{Sahoo2021}. Surprisingly, from the figure, it can be seen that there are scarcely a few cases available at high $M_\infty$ and high $Re_D$.} 

\sk{As mentioned earlier, the unsteady pulsation flow modes have been studied by researchers in the past both experimentally and numerically \citep{Ericsson1976,Reding1977,Geng200685,Mehta2002,Mehta2002431}. Several theories have been proposed concerning the driving mechanism for these flows \citep{Karlovskii1986437,Khlebnikov1996731,Mair1952,Maull1960584}. Feszty \etal \citep{Feszty2004a} carried out a numerical investigation, studying the laminar flowfield and the driving mechanism for pulsation over a spiked cylinder configuration with $[l/D] = 1$. They identified the following processes to occur in a pulsation cycle, namely collapse, inflation, and withhold, based on which the driving mechanism was explained in detail. A vortical region is formed near the cylinder-spike junction during the collapse stage. It is the high-pressure gas that was trapped in this vortical region expanding violently during the inflation stage, rather than the continuous mass influx due to Edney's type-IV interaction \cite{Edney1968}. These features are identified as the driving mechanism for the self-sustained shock motion called pulsation, unlike the other means that were previously thought by the other researchers\cite{Mair1952,Maull1960584,Antonov1976746,Panaras200969}. Even some of the results from the laminar flow numerical investigations were found to agree with the experimental work (flowfield visualization) of Kenworthy\citep{Kenworthy1978}.} 

\sk{However, our recent experimental campaign on pulsating flowfields, especially over a spiked-blunt-forebody of base diameter $D$ at high $Re_D$ and freestream $M_\infty$ revealed exciting results, especially during the collapse phase of pulsation, which made us re-think the driving mechanism behind pulsation. The motivations behind our current experimental campaign are two folds: 1. At high $Re_D$, the turbulent flow effects might be strong, which is different from the laminar observation done by Feszty \etal \citep{Feszty2004a}. In fact, from the brief literature review, the authors found that the research works at high $Re_D$ and high $M_\infty$ on spiked body flows are scarcely available (see Figure \ref{fig:literature_cases}); 2. High-speed schlieren imaging and pressure measurements at hypersonic Mach numbers will shed valuable information on the formation of typical flow structures that reveal the alterations in the pressure loading on the vehicle itself. If a clear understanding of the flowfield mentioned above is attempted, then only formulating an efficient active or passive control device would be feasible.}

\sk{With the motivations mentioned above, the authors took the following as their distinct objectives for the present work:
\begin{enumerate}
    \item{To experimentally study the pulsating flowfield over a cylindrical forebody with a spike of $[l/D]=1$, in a hypersonic flow with $M_\infty= 8.1$ at two different $Re_D$.}
    \item{To obtain the pressure distribution on the flat-face cylindrical forebody and to visualize the flowfield using high-speed Schlieren imaging.}
    \item{To compute and compare the variations in the pressure loading, fluctuation intensity, and spectral decay from the unsteady pressure measurements.}
    \item{To understand the driving flow modes from the high-speed schlieren images after subjecting them through modal analysis (Proper Orthogonal Decomposition-POD and Dynamic Mode Decomposition-DMD).}
\end{enumerate}}

The rest of the paper is organized as follows: Details about the experimental methodology is given in $\S$-\ref{sec:experimentation} and followed by the experimental uncertainties in $\S$-\ref{sec:uncertainty}. Results and discussions are given in $\S$-\ref{sec:res_disc} under different subsections: high-speed schlieren images at $\S$-\ref{sec:high_speed_img}, $x-t$ diagram at $\S$-\ref{sec:x-t_diag}, unsteady pressure signals at $\S$-\ref{sec:unsteady_p}, and the modal analysis at $\S$-\ref{sec:modal_analysis}. Some of the major conclusions are presented in $\S$-\ref{sec:conclusion}.

\begin{figure*}
	\centering \includegraphics[width=0.8\textwidth]{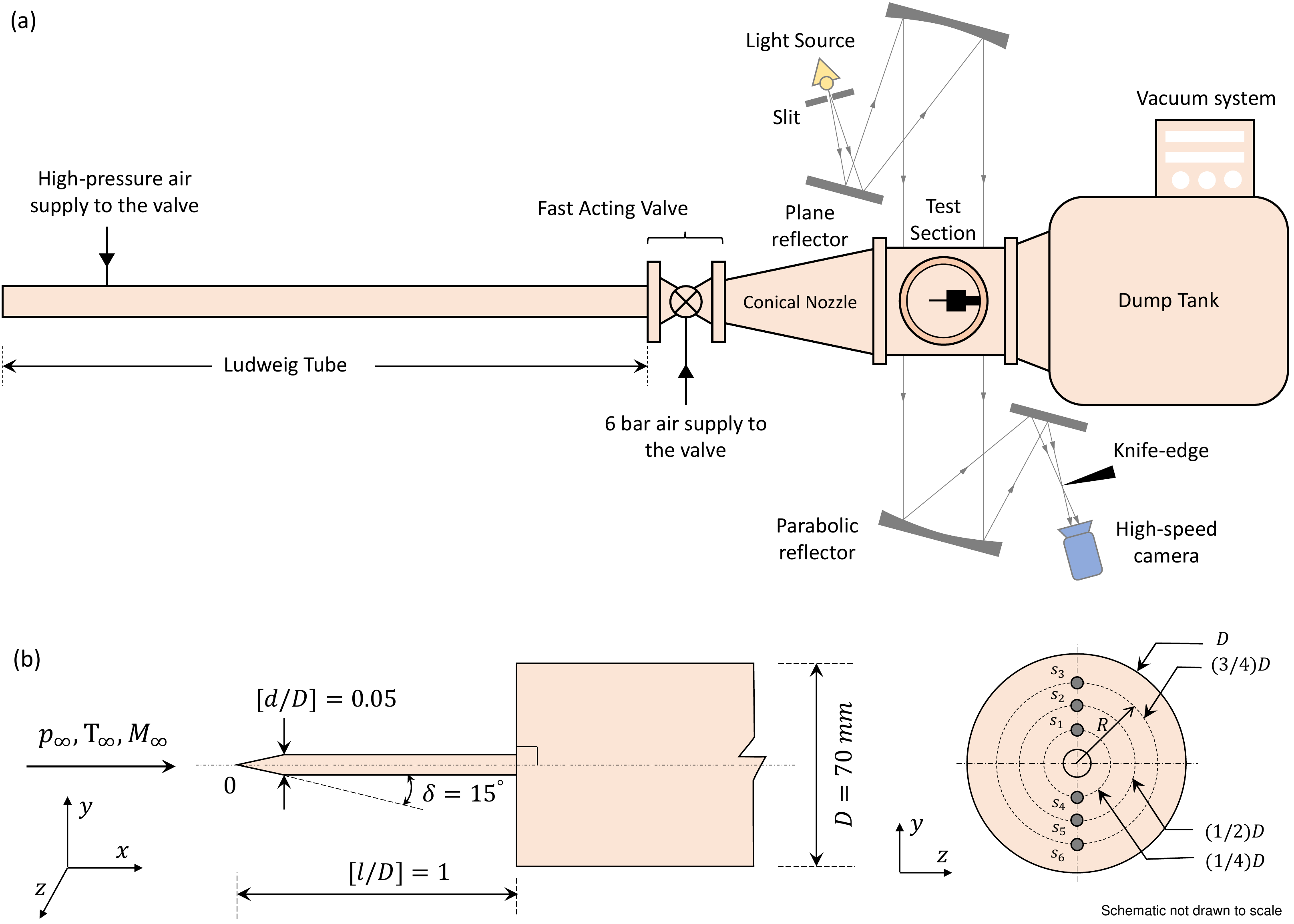}
	\caption{\sk{(a). A schematic (not drawn to scale) showing the top view of the Ludwig facility at IISc-Bengaluru with the `Z-type' schlieren arrangement\cite{Settles2001} to study the pulsating flowfield observed around an axisymmetric flat-face spiked body in a hypersonic flowfield; (b). A schematic (not drawn to scale) showing the geometrical configuration of the model under investigation and the locations of the unsteady pressure transducers ($S_1-S_6$) mounted at different radial locals ($R/D$). Model is oriented against the flow direction.}}
	\label{figure1}
\end{figure*}
 
\section{Experimentation} \label{sec:experimentation}
\subsection{Test Facility} \label{sec:test_facility}
Experiments are performed in the recently developed IISc Ludwig tunnel, and the schematic (not drawn to scale) is shown in Figure \ref{figure1}a. Similarly, details about the testing model are given in Figure \ref{figure1}b. More details about the test model are given in $\S$-\ref{sec:instruments}. The Ludwieg tunnel is a modification of the existing hypersonic shock tunnel: HST-2 \citep{Sriram2016}, by merging the driver and driven sections to form a Ludwig tube. It is essentially a pressure tube where the test gas (nitrogen, $N_2$) is filled to the required pressure ($p_{01}$). The end of the Ludwig tube (length 7.12 m and inner diameter 50 mm) is connected to a commercially available fast-acting valve, ISTA\textsuperscript{\textregistered} KB-40. It is a pneumatically assisted solenoid valve, requiring 6 bar pressure and 24 V DC for its operation. The valve response time for the input volt pressure is around 1 ms. The valve isolates the dump tank, containing the nozzle and test section at an ultra-low vacuum of $10^{-8}$ bar.

\begin{table*}
\caption{\sk{Freestream flow conditions achieved in the test section of the Ludwieg tunnel during the present investigation for two different freestream Reynolds number ($Re_D$) cases based on the base body diameter ($D$) at a constant freestream Mach number of $M_\infty=8.16$.}}
	\begin{ruledtabular} 
		\begin{tabular}{lll}
		    \textbf{Quantities} & \textbf{Case - A} (low $Re_D$) & \textbf{Case - B} (high $Re_D$)\\
		    \midrule
			Total Pressure ($p_{01}$, Pa)  &
			$10 \times 10{^{5}} {\pm} 5\%$ & $40 \times 10{^{5}} {\pm} 5\%$\\
			Total Temperature ($T_{01}$, K)  &
			$300 {\pm} 2\% $ & $300 {\pm} 2\% $\\
			Freestream Pressure ($p_\infty$, Pa)  &
			$90.05	{\pm} 5\%$ & $360.19	{\pm} 5\%$\\
			Freestream Temperature ($T_\infty$, K)  &
			$20.95 {\pm} 2\%$ & $20.95 {\pm} 2\%$\\
			Freestream Density ($\rho_\infty$, kg/m$^3$)  &
			$0.02 {\pm} 5\%$ & $0.06 {\pm} 5\%$\\
			Freestream Velocity ($u_\infty$, m/s)  &
			$748.66	{\pm} 2\%$ & $748.66	{\pm} 2\%$\\
			Freestream Kinematic Viscosity ($\nu_\infty$, m$^2$/s) &
			$6.87 \times 10{^{-5}} {\pm} 2\% $ & $1.72 \times 10{^{-5}} {\pm} 2\% $\\
			Freestream Mach number ($M_\infty$) &
			$8.16 {\pm}1\%$ & $8.16 {\pm}1\%$\\
			Reynolds number based on $D=70$ mm ($Re_D=u_\infty D/\nu_\infty$) &
			$0.76 \times 10{^{6}} {\pm} 5\%$ & $3.05 \times 10{^{6}} {\pm} 5\%$\\
			Equivalent Altitude ($h$, km) & 21.8 & 13.2 \\
		\end{tabular}
		\label{table1}
	\end{ruledtabular}
\end{table*}

The operation of the valve results in the expansion of high-pressure gas in the pressure tube into the test section of cross-section 300 mm $\times$ 300 mm and a length of 450 mm while passing through the diverging nozzle of design Mach number $M_{\infty,D}=8$ (see Figure \ref{figure1}a). Later, the flow exits into the 2.5 m long dump tank. The facility is designed to generate the desired freestream flow with a uniform core flow diameter of 240 mm and unit Reynolds number varying from 10 – 90 $\times10^6$ /m depending on the fill pressure. \sk{The present experiments consider two different fill pressures to simulate two different $Re_D$: a. 10 bar for low a $Re_D$ of $0.76 \times 10^6$, and b. 40 bar for a $Re_D$ of $3.01 \times 10^6$. The respective $Re_D$ are selected in a manner to simulate the flight conditions at altitudes of 22 km and 13 km, where the spiked body flows have relevance in terms of axisymmetric scramjet inlet operation \cite{oswatitsch1957,lu1998} and ballistic missile drag reduction \cite{Mair1952,bogdonoff1959,Maull1960584,Yamauchi1995}.} 

\begin{figure*}
	\centering \includegraphics[width=0.8\textwidth]{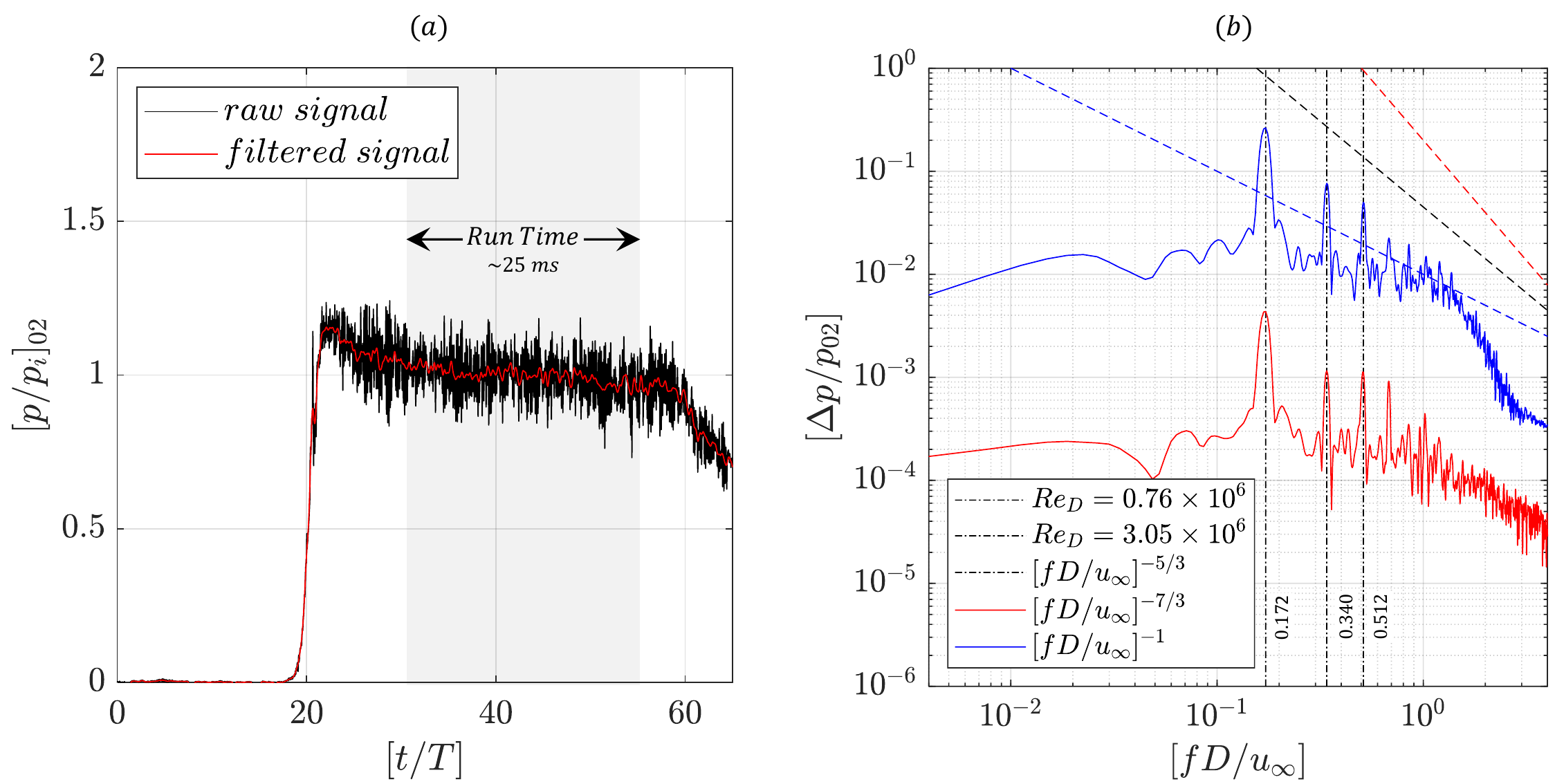}
	\caption{(a). A typical pitot pressure signal ($p_{02}$) shows a region of almost constant run-time from the mounted pitot tube in the Ludwig tunnel's test section. The $x$-axis is normalized with $T=1$ ms and the $y$-axis is normalized with the ideal or isentropic pitot-pressure ($p_{02,i}$) based on the design Mach number ($M_{\infty,D}$); \sk{(b) Normalized pressure amplitude spectra ($\Delta p/p_{02}$) showing the dimensionless spectral decay ($fD/u_\infty$) observed from the unsteady pressure signal at $s_4$ sensor location for two different $Re_D$ cases. The decay rate (dash lines) is also compared with the inverse law, -7/3 law, and -5/3 law for evaluating the behavior of turbulent structures formed during the pulsation events.}}
	\label{figure2}
\end{figure*}

The Pitot pressure ($p_{02}$) is measured simultaneously along with the test model during the experiments in order to ascertain the freestream conditions (see Table \ref{table1}). The achieved test time during a typical high $Re_D$ case is shown in Figure \ref{figure2}a. A steady flow test period of around 25 ms was observed for the reported experiment, as seen from the Pitot pressure signal. \sk{In the present set of experiments, the facility was operated to give two different Reynolds numbers of 11 $\times 10^6$/m and 43$\times10^6$ /m, by filling the Ludwig tube with nitrogen at $p_{01}=10$ bar and $p_{01}=40$ bar, respectively. A typical normalized pressure amplitude spectral decay observed from the $s_4$ unsteady pressure sensor is plotted in Figure \ref{figure2}b for two different $Re_D$ cases. The dimensionless frequency is given by $[fD/u_\infty]$, where the freestream velocity as tabulated in Table \ref{table1} remains constant for both the cases. Initial observation shows that the dominant components remain invariant with $Re_D$. More details on the unsteady pressure spectra are briefly discussed in the upcoming sections, particularly at $\S$-\ref{sec:unsteady_p}.} 

\subsection{Test Model, Instrumentation and Flow Diagnostics} \label{sec:instruments}
The test model used is a flat-face cylinder with a diameter of $D=70$ mm, having a forward-facing spike of length $l=70$ mm ($l/D=1$, fineness ratio) with a conical tip having a semi-apex angle of $\delta = 15^\circ$. The spike has a slenderness ratio of $[d/D]=0.05$, \sk{where $d$ is the diameter of the spike}. The model is instrumented with high-frequency pressure transducers (PCB Piezotronics) of 1 MHz acquisition rate flush-mounted on its flat-face at radial ($R$) distances of 17.5 ($D/4$), 35 ($D/2$), and 52.5 ($3D/4$) mm from the axis along the transverse direction, to measure the unsteady pressure fluctuations. A schematic of the test model and the sensor mounting location is shown in Figure \ref{figure1}b.

\begin{figure}
	\centering \includegraphics[width=\columnwidth]{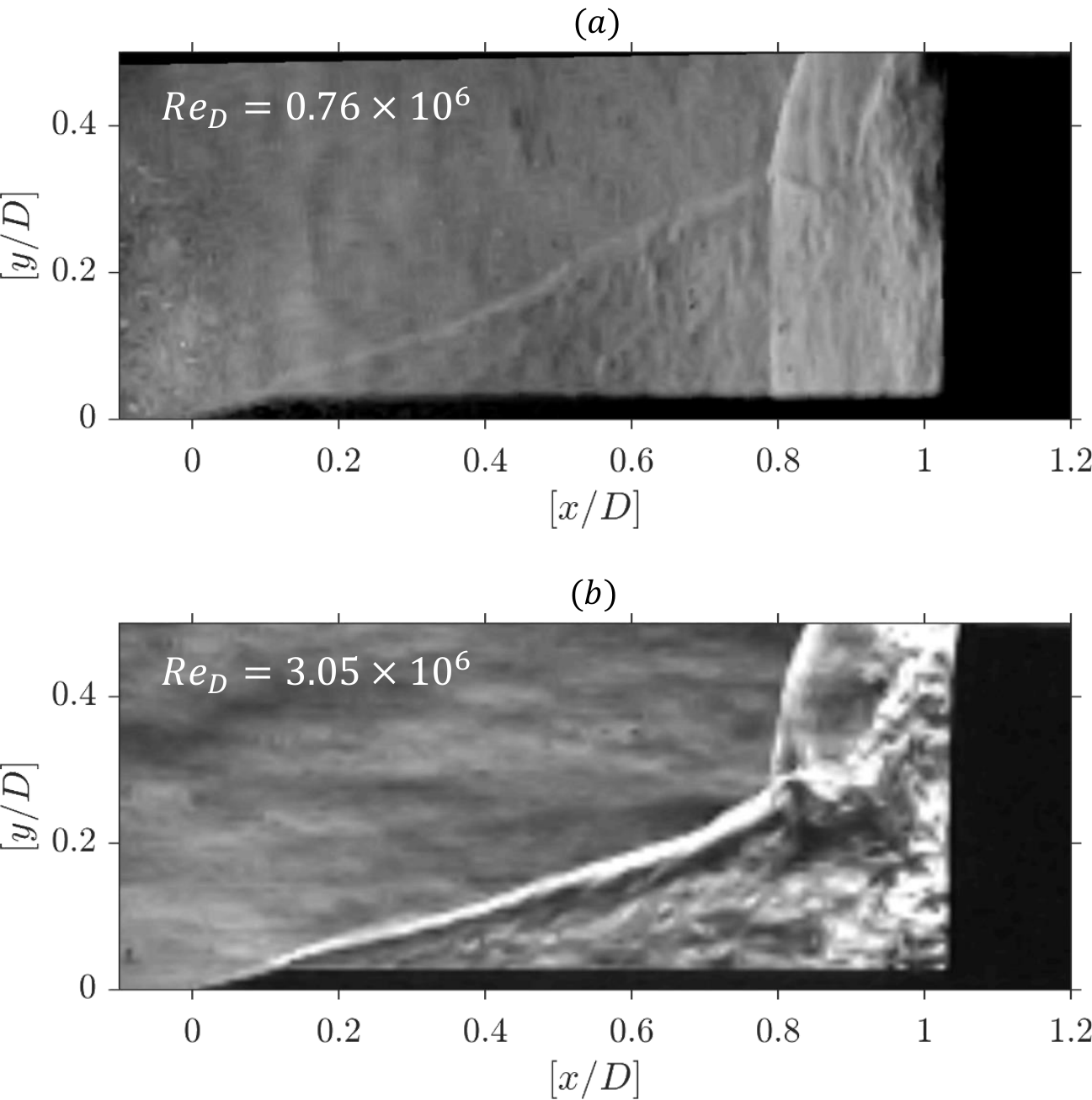}
	\caption{\href{https://youtu.be/kR7Rps5POWk}{(Multimedia View)} \sk{A typical instantaneous schlieren image depicting the line-of-sight integrated normalized density gradients in $y$-direction ($\left \Vert (\partial \rho/\partial y\right) \Vert$) taken at an arbitrary time interval ($\tau$) for two different freestream Reynolds number based on the base body diameter ($Re_D$) at a freestream Mach number of $m_\infty=8.16$: (a). $Re_D = 0.76 \times 10^6$ (low $Re_D$, case-A), and (b). $Re_D = 3.01 \times 10^6$ (high $Re_D$, case-B). Detailed flow features are marked in Figure \ref{figure4} as they are almost common for both cases.}}
	\label{figure3}
\end{figure}

The hypersonic pulsating flowfield over the test model was visualized using `Z-type' high-speed schlieren technique \citep{Settles2001} using a Phantom V310 high-speed camera, operated at 40 kHz, with an exposure time of 2 $\mu$s. The frame size was 256 $\times$ 256 pixels with a pixel resolution of $\sim 0.45$ mm/pixel. \sk{A typical instantaneous schlieren image showing the normalized line-of-sight integrated density gradients in $y$-direction is shown in Figure \ref{figure3} for two different $Re_D$ cases. The sensitivity of schlieren imaging plays a vital role in resolving the flow features. Therefore, flow features are not captured with good contrast in Figure \ref{figure3}a owing to the low density, but they are in Figure \ref{figure3}b.} The schlieren setup, as shown in Figure \ref{figure1}a utilizes a light source made from an array of white LEDs of 7 W power after passing through a pin-hole slit. The knife-edge was kept horizontal at the point of focus, enabling us to see the light intensity changes due to density gradients in the vertical direction ($\partial \rho/\partial y$), as the flow features (shock-shock interaction, separation region, and vortical regions, see Figure \ref{figure4}) were visualized only in this orientation of the knife-edge.

\section{Measurement Uncertainties}\label{sec:uncertainty}
All the experiments were repeated at least three times to ensure statistical consistency and repeatability. Steady and unsteady measurements suffer from uncertainty due to repeatability, acquisition, data conversion, storage, and sensitivity to external factors. The steps are given in the text of Coleman and Steele\cite{Coleman2009} and the recommendations while taking pitot measurements in Sutcliffe, and Morgan \cite{Sutcliffe_2001} were followed to compute the total uncertainty in pressure data. Uncertainties from the images and the derived data were computed using the principles given by Santo \etal \cite{Santo2004}. All the signals were preconditioned, and no further post-processing was done, including padding, windowing, or spectral smoothing. Measurements of unsteady pressure and schlieren imaging were done simultaneously to resolve and associate the flow features with the respective dynamic events. All the steady-state or low-response pressure transducers like the driver pressure monitor or $p_{01}$ include a total uncertainty of $\pm 5\%$ about the mean. \sk{The unsteady pressure transducers also exhibit a total uncertainty of $\pm 5\%$ about the mean with a spectral resolution of $\Delta f_p\approx 40$ Hz (the total number of considered samples ($n$) are 25001, and the sampling rate ($f_s$) is 1 MHz which results in the spectral resolution of $\Delta f=f_s/n=39.9984$ Hz).} The schlieren images and the resulting spatial modes from the modal decomposition exhibit a spatial resolution of $\Delta x\approx \Delta y\approx 0.5$ mm. The spectral resolution from the modal decomposition is $\Delta f_m \approx 40$ Hz.   

\section{Results and Discussions}\label{sec:res_disc}

\begin{figure*}
	\centering \includegraphics[width=0.7\textwidth]{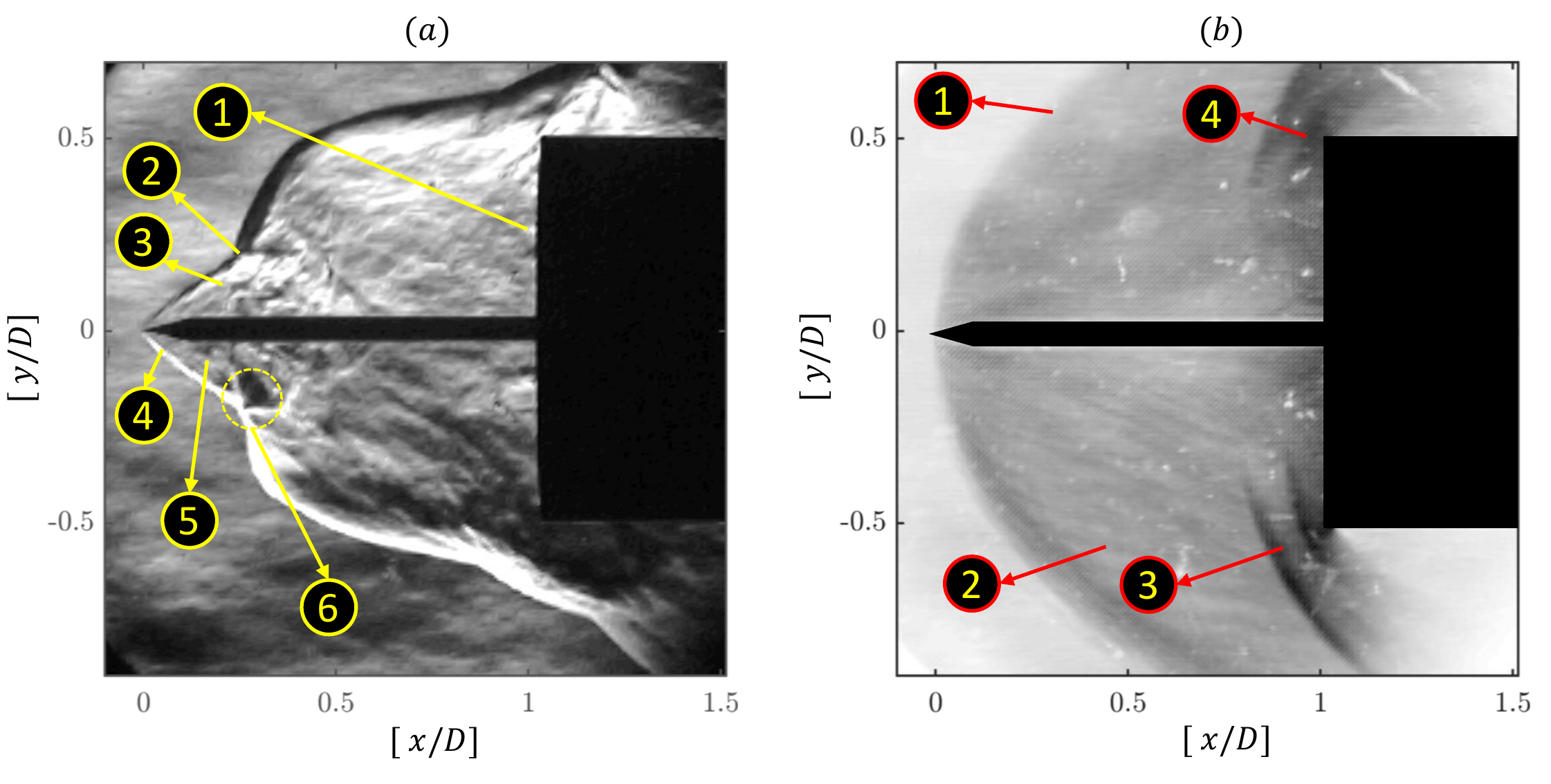}
	\caption{(a) \href{https://youtu.be/Hm8VkG-xZUY}{(Multimedia View)} Instantaneous schlieren imaging showing the pulsating flowfield at an instant of the `collapse' phase for case-B (high $Re_D$); Dominant flow features: 1. gas compression due to collapse in front of the flat-face, 2. triple point ($\lambda$-shock), 3. separated free shear layer, 4. leading-edge shock, 5. recirculation zone, 6. shedding vortices from the triple point. (b) Operator based time-averaged image ($\left\|\bar{\boldsymbol I}-\boldsymbol I_{rms}\right\|$) showing the extent of pulsating shock-laden flowfield. Dominant flow features: 1. inflated shock in the leading edge of the spike; 2. trace of the separation shock during the time of inflation; 3. Collapsed shock forming ahead of the flat-face; 4. Shock formation due to the rapid compression of collapsing flowfield.}
	\label{figure4}
\end{figure*}

\subsection{Flow physics from high-speed schlieren images}\label{sec:high_speed_img}
A pulsating flowfield around the spiked body has many flow features. High-speed schlieren helps in identifying them. A typical instantaneous and time-averaged schlieren image is given in Figure \ref{figure4}. As low $Re_D$ case results in low density and the sensitivity of the schlieren imaging system has limitations in resolving small density gradients, hereafter in the discussions, only high $Re_D$ case is considered unless otherwise specified. Some of the flow features, including separated free shear layer, vortices from the triple point, and recirculation region, can be seen from the instantaneous image (Figure \ref{figure4}a). The vortical structures are identified with surety because previously, in Feszty \etal \cite{Feszty2004a}, the authors matched the vortical zones from their computations with that of the schlieren images taken by Kenworthy\cite{Kenworthy1978}. The comparison further revealed the locations where the vortical zones are expected in the schlieren images.

One of the dominant flow features like the moving shocks from the flat-face to the spike-tipis captured in Figure \ref{figure4}b where the operator based time-averaged image ($\left\|\bar{\boldsymbol I}-\boldsymbol I_{rms}\right\|$) is shown. The presence of two strong shocks in front of the forebody is vital to note, as it is the characteristics observed at higher $Re_D$ and $M_\infty$, which was not observed before. In addition, the shock angle from the leading edge of the spike-tip is shallower at higher freestream hypersonic Mach numbers than that of the lower supersonic freestream Mach numbers. Shallow shock angle meets the forebody halfway and impinges on the solid forebody resulting in the formation of Edney's type-IV\cite{Edney1968} shock interference pattern. The associated flow events have similarities with relevant pulsating flowfields at other speed regimes; however, they are not identical \citep{Mair1952,Maull1960584,Kenworthy1978,Feszty2004a}. Respective discussions towards those non-identical flow events are explained further.

\begin{figure*}
	\centering \includegraphics[width=1\textwidth]{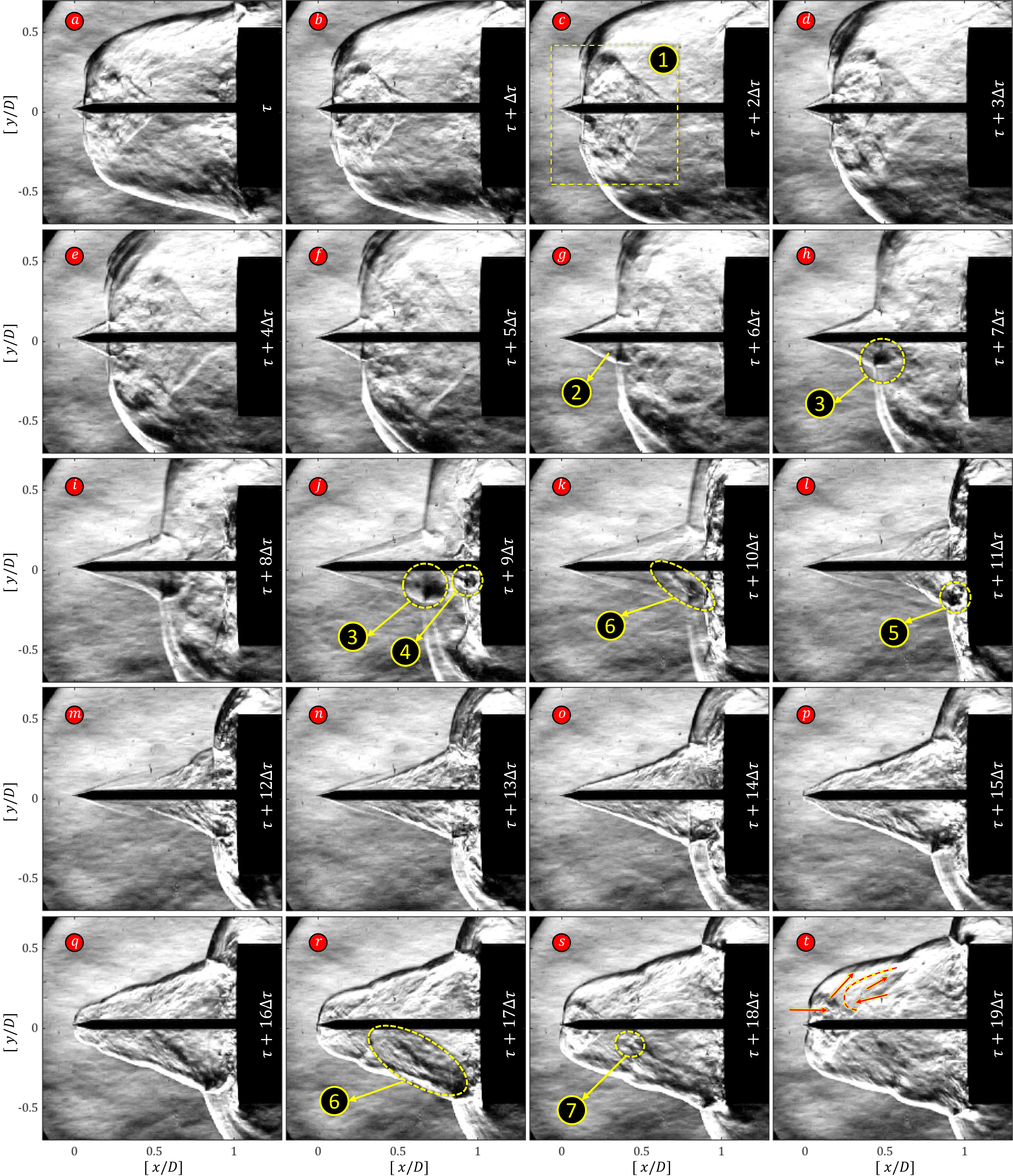}
	\caption{\href{https://youtu.be/Hm8VkG-xZUY}{(Multimedia View)} Instantaneous schlieren image frames representing a pulsating flow cycle from a typical time of $\tau$ to $\tau+19\Delta \tau$ where $\Delta \tau=25$ $\mu$s for a high $Re_D$ case. Flow is from left to right. Dominant flow features:  1. Beetle-leaf-like structure, 2. Separation shock, 3. Vortical region (V1) from the triple point of $\lambda_1$-shock, 4. Vortical region (V2) from the triple point of $\lambda_2$-shock, 5. Merging of vortical regions, 6. Supersonic jet with shock-cells, 7. normal shock from the impinging supersonic jet at an oblique angle.}
	\label{figure5}
\end{figure*}

\begin{figure*}
	\centering \includegraphics[width=0.7\textwidth]{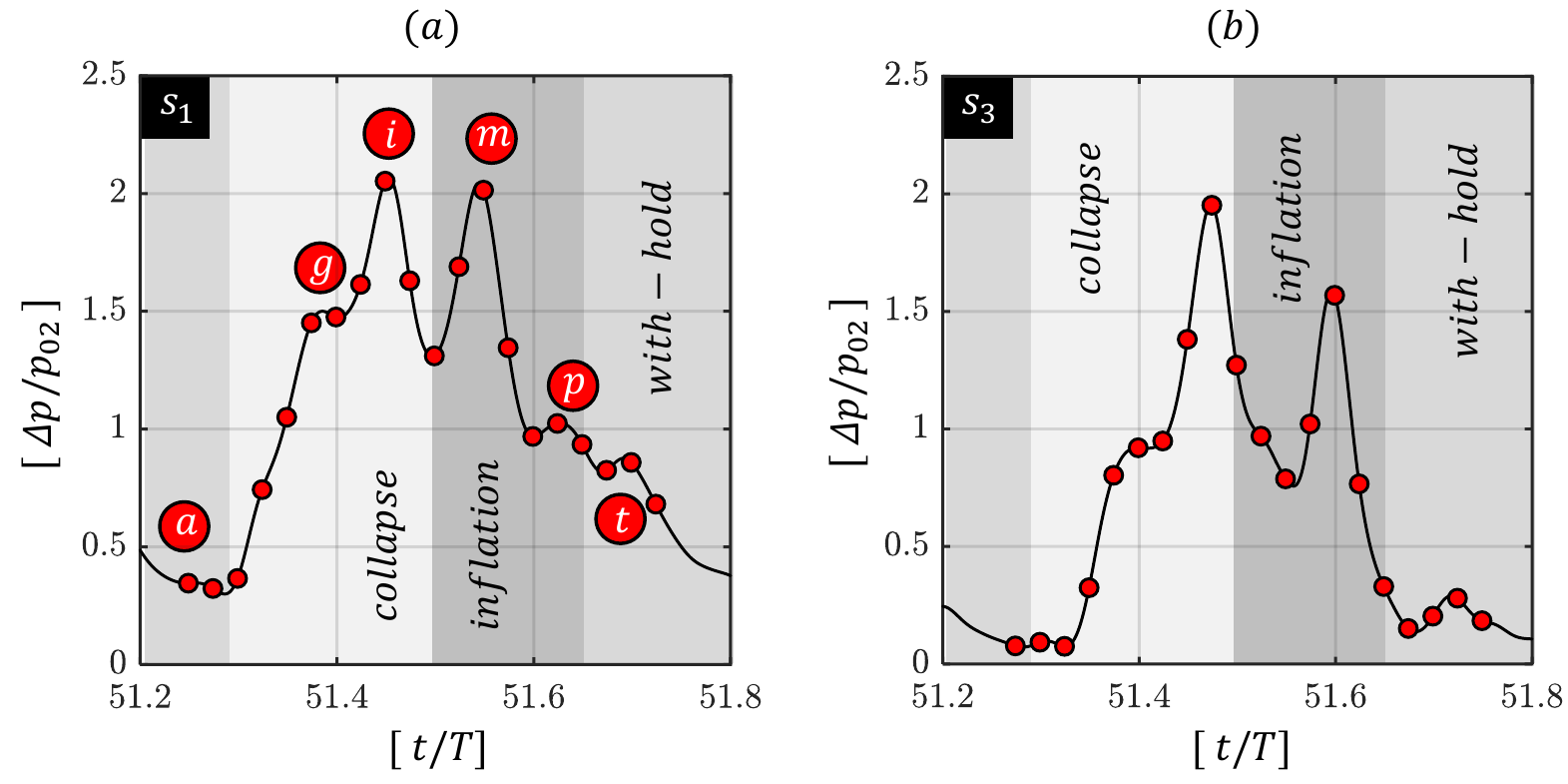}
	\caption{A typical normalized pressure cycle ($\Delta p/p_{02}$) observed at (a) $s_1$ and (b) $s_3$ sensor location during a typical run-time for a high $Re_D$ case. The red color-filled circle and the corresponding text markers represent the event location in accordance with the instantaneous schlieren images available (from $\tau$ to $\tau+19\Delta \tau$) in Figure \ref{figure5}. The shaded regions demarcate the different regimes of a pulsation cycle.}
	\label{figure6}
\end{figure*}

A pulsation cycle consist of three stages. All the three stages are shown in Figure \ref{figure5} as sequence of images, starting from $\tau = 0$ to $\tau + 19\Delta \tau = 475$ $\mu$s with $\Delta \tau = 25$ $\mu$s. The start to end of a pulsation cycle (frames \textcircled{a} - \textcircled{t}) and the events corresponding to the observations are also marked on the obtained pressure signals, near root ($s_1=0.25D$) and shoulder ($s_3=0.75D$) region of the test model, as shown in Figure \ref{figure6}a and \ref{figure6}b. 

`Collapse' stage (marked between frame \textcircled{a} at $\tau$ to frame \textcircled{h} at $\tau+7\Delta \tau$) is characterized by unsteady shock motion from the tip of the spike towards the after body. `Inflation' stage (marked between frame \textcircled{i} at $\tau+8\Delta \tau$ to frame \textcircled{p} at $\tau+15\Delta \tau$) follows next, where the high pressure gas trapped during the collapse stage expands rapidly in the recirculation region. `With-hold' (marked between frame \textcircled{q} at $\tau+16\Delta \tau$ to frame \textcircled{t} at $\tau+19\Delta \tau$) is the last stage where the shock remains stationary and the exploding high pressure gas escapes through the shoulder of the cylindrical after body and initiates a `collapse' cycle \citep{Feszty2004a}. 

Our discussion focuses mainly on the `collapse' stage of a pulsation cycle. During this stage, the unsteady shock-shock interaction has led to the formation and merging of toroidal vortices ($\tau+9\Delta \tau$), rapidly growing in size and contributing to a sustaining event of pulsation. 

The cycle starts with the interaction of oblique shock emanating from the tip of the spike with the bow body shock wave as seen in \textcircled{a} at $\tau$. The shock system described above, named Converging Shock System (CCS), is unsteady and starts moving towards the right, as can be seen in the subsequent frames (till frame \textcircled{p} at $\tau+15\Delta \tau$). The shock interaction, mainly seen in frame \textcircled{h} at $\tau+7\Delta \tau$, is similar to the pattern described by Edney\citep{Edney1968}. However, it is unclear whether it is a \sk{type-III or type-IV} interaction\citep{Edney1968} from the present flow visualization. At the point of interaction (from frame \textcircled{h} at $\tau+7\Delta \tau$ to frame \textcircled{j} at $\tau+9\Delta \tau$), \sk{i. e. triple point, a shear layer exists, due to the velocity gradient\cite{gatski2013,gnani2014}, forming the boundary between the subsonic flow behind the bow shock and supersonic flow behind the reflected oblique shock wave.} The reflected oblique shock wave impinges on the boundary layer of the spike’s body surface, causing flow separation, which shall be discussed in the subsequent paragraph.

At the same time, there is a rapid expansion of high-pressure gas at supersonic speeds \citep{Kariovskii1987601}, in the opposite direction to the motion of the shock system, corresponding to the inflation stage of the previous pulsation cycle (supersonic jet with shock-cells in frame \textcircled{a} at $\tau$). At the time of `collapse,' there is a clear demarcation between the two flowfields as can be very clearly seen in frame \textcircled{c} at $\tau+2\Delta \tau$ in the form of a beetle leaf-like structure. This demarcation line/boundary is a shock wave separating the supersonic flow on the side (cylindrical afterbody side) to the mixed subsonic/supersonic flow on the other side (spike tip side). With time, the beetle leaf-like structure expands in size (from \textcircled{d} at $\tau+3\Delta \tau$ to \textcircled{f} at $\tau+5\Delta \tau$). The beetle leaf-like structure then runs along the spike and then impinges on the cylindrical afterbody as seen in frame \textcircled{g} at $\tau+6\Delta \tau$ and moves laterally along its surface, as seen in subsequent frames \textcircled{g} at $\tau+6\Delta \tau$ and \textcircled{h} at $\tau+7\Delta \tau$.   

The shock wave impinges on the forebody causing peak pressure loads on the forebody (frame \textcircled{l} at $\tau+11\Delta \tau$). The shock wave then moves over the shoulder, exposing the aft body region near the root of the spike to supersonic flow. The phenomenon mentioned above explains the formation of normal shock wave ahead of the cylindrical afterbody, its interaction with the spike boundary layer, leading to flow separation and finally forming a toroidal vortical region near the root of the spike (frame \textcircled{h} at $\tau+7\Delta \tau$), as explained by Feszty \etal \citep{Feszty2004a} in his work. A similar phenomenon was also observed in our experimental investigation, as can be seen in frame \textcircled{j} at $\tau+9\Delta \tau$. Here the vortical region is numbered 3 (V1) and 4 (V2). \sk{Vortical region-V1 comes from the triple point of the lambda-$\lambda$ shock formed from the spike-tip. On the other hand, vortical region-V2 forms from the triple point of the lambda-$\lambda$ shock forming closer to the blunt-body surface. The formation of V2 is purely due to the `collapse' phase, as the compressed gas during collapse accumulates in front of the forebody surface and generates another shock that travels upstream. Feszty \etal \citep{Feszty2004a} believed the vortical region (V2) to be the driving mechanism for pulsation.}

Meanwhile, as the CSS advances towards the aft body, the reflected shock emanating from the triple point, as seen in frame  \textcircled{g} at $\tau+6\Delta \tau$, impinges on the spike body surface and causes the boundary layer to separate, resulting in the formation of a separation bubble and shear layer over it. As the shock system moves downstream, the separation bubble and shear layer grows in size (from \textcircled{h} at $\tau+6\Delta \tau$ to frame  \textcircled{k} at $\tau+10\Delta \tau$). As already discussed, at the triple point, due to velocity gradient, a shear layer exists. Since the pressure behind the bow shock is relatively high compared to the pressure behind the oblique shock wave, the shear layer curls inwards (towards the spike), forming a toroidal vortical region V1, frame \textcircled{g} at $\tau+6\Delta \tau$ of Figure \ref{figure5}. This vortical region is continuously fed from the growing shear layer (S2), resulting in increase in its size, frame \textcircled{g} to \textcircled{k} at $\tau+10\Delta \tau$ of Figure \ref{figure5}. A significant mass (high density) is trapped inside this vortical region V1. As the shock system moves close to the cylindrical afterbody, the vortical regions V1 and V2 interact with each other and the cylindrical afterbody, resulting in its breakdown and release of the trapped high-density gas within it.

Unlike the earlier findings by Feszty \etal \citep{Feszty2004a}, where it was observed that the gas trapped inside the vertical region V2, causing the rapid expansion, it is clear that it is not just the vortical region V2 near the root of the spike but also the growing vertical region V1, which contributes to the rapid expansion. The interaction of the votrical regions (V1 and V2 and with the wall) followed by the rapid release of the trapped gas inside them results in pressure rise near the root of the spike as seen in Figure \ref{figure6}a, marked \textcircled{l} at $\tau+11\Delta \tau$. \sk{It has to be emphasized here that V1 is formed from the triple point of the $\lambda$-shock system arising due to the impingement of the shallow oblique shock on a solid boundary (base body). Similarly, V2 is formed from building another $\lambda$-shock system and its corresponding triple point ahead of the base body. A proof of vortices shedding from a $\lambda$-shock's triple point is shown in a \href{https://youtu.be/dEvM51VeDCw}{video} available in the supplementary for a similar $Re_D$ and $[l/D]$ as evidence through a Detached Eddy Simulation (DES) whose details are beyond the scope of present investigations.}

After the release, the trapped gas expands in the lateral and axial direction, starting from frame \textcircled{m} at $\tau+12\Delta \tau$ till the end of the cycle, marked by pressure drop on the cylindrical face, as can be seen in Figure \ref{figure6}a. As it expands in the lateral direction, the aft body shock approaches close to the cylindrical body, resulting in pressure rise near the shoulder of it as marked \textcircled{m} at $\tau+12\Delta \tau$ in Figure \ref{figure6}b, followed by its drop as the gas expands.
The expanding gas towards the upstream direction, named expanding shock system (ESS), is bounded by oblique shock and bow shock wave with a shear layer at their point of interaction, and its first occurrence is seen clearly in frame \textcircled{n} at $\tau+13\Delta \tau$ of Figure \ref{figure5}. One should not confuse this shock wave with the oblique shock wave emanating from the spike tip. With the rapid expansion of the gas, this shock system moves in axial (away from aft body) and lateral direction. The shear layer impinges on the cylindrical afterbody resulting in pressure rise on its surface as \sk{can be seen} in Figure \ref{figure6}a marked \textcircled{n} at $\tau+13\Delta \tau$ and in Figure \ref{figure6}b marked \textcircled{r} at $\tau+17\Delta \tau$. The shock system reaches out to the tip of the spike, where it changes its shape from oblique to bow shock (frame \textcircled{r} at $\tau+17\Delta \tau$). The ESS remains attached to the tip of the spike, and expansion happens only in the lateral direction, which is the withhold phase of the cycle. The sequence of events mentioned above continues to occur again in the next pulsation cycle. 

\subsection{Shock foot-print analysis}\label{sec:x-t_diag}

\begin{figure*}
	\centering \includegraphics[width=1\textwidth]{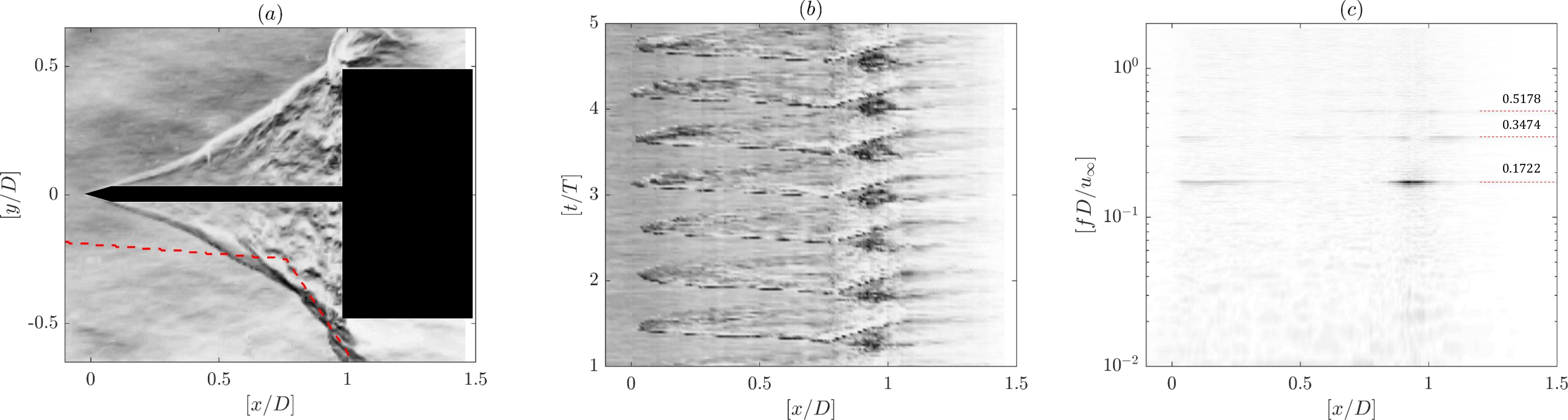}
	\caption{\href{https://youtu.be/zBK31t6TN38}{(Multimedia View)} (a) Extracting the shock foot trajectory by grabbing the image light intensity along the dotted red-line from every schlieren image for the high $Re_D$ case; (b) Constructed $x-t$ diagram shown for a typical normalized time scale showing the periodic shock motion ($T=1$ ms); (c) Intensity normalized Fast-Fourier Transform of the $x-t$ diagram showing the presence of discrete dimensionless frequencies as similar to that of the unsteady pressure signal in Figure \ref{figure2}b.}
	\label{figure7}
\end{figure*}

The sequence of shock motion and the associated flow physics can be explained through a $x-t$ diagram (see Figure \ref{figure7}). \sk{A suitable line profile along the flow direction is first drawn to begin constructing the $x-t$ diagram (Figure \ref{figure7}a). The line is drawn so that the oscillating shock's path or footprint is passing through the line. Such a passing renders the $x-t$ diagram over a while by registering the shock's trace with good contrast (Figure \ref{figure7}b). The cause mentioned above is the primary reason for not picking a simple straight line parallel to the spike-stem}. As described earlier, in Figure \ref{figure7}a the line segment along which the $x-t$ diagram is constructed has been given in dotted red-line. The intensity variations along that line in each of the captured images are piled upon to trace the shock motion as shown in Figure \ref{figure7}b. A periodic shock oscillation of 10-cycles can be seen in it. \sk{By performing a fast Fourier transform (FFT) along the $y$-axis for all the $[x/D]$ locations, a $x-f$ diagram is constructed, where the dominant frequencies can be observed as shown in Figure \ref{figure7}c. The capturing of these frequencies will not be possible if the line segments in Figure \ref{figure7}a are not selected properly. Both visual cues from the high-speed schlieren imaging and the FFT of the unsteady pressure signals thus, help in constructing an appropriate $x-t$ and $x-f$ diagrams.} The fundamental frequency ($[f_1D/u_\infty$) is observed at about 0.172 and two clear overtones are seen at $[f_2D/u_\infty]\approx 0.3432$ and $f_3\approx 0.5189$. They exist in a relationship of $[f_3D/u_\infty]=3[f_1D/u_\infty]$ and $[f_2D/u_\infty]=2[f_1D/u_\infty]$, which indicates a self-sustained harmonic behavior of the oscillating shock systems.

\begin{figure*}
	\centering \includegraphics[width=0.8\textwidth]{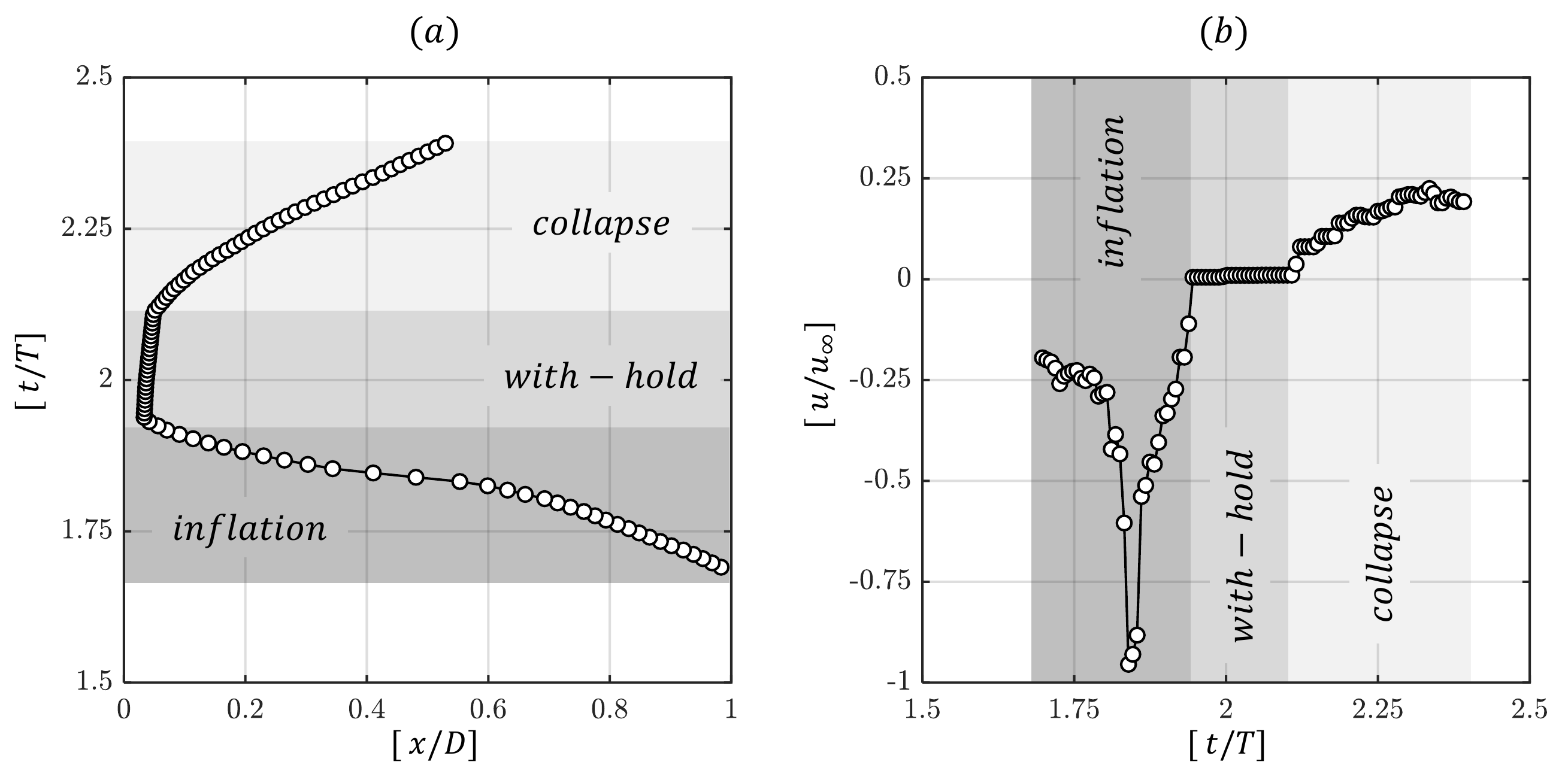}
	\caption{(a) A typical $x-t$ trace from a cycle of pulsating shock-motion shown in Figure \ref{figure7} for a high $Re_D$ case, explaining the presence of distinct three cycles: inflation, with-hold, and collapse; (b) Plot showing the variation of shock speed at different time instants during the considered pulsation cycle by measuring the slope along the shock trajectory.}
	\label{figure8}
\end{figure*}

In addition to the temporal details about the shock motion, the $x-t$ diagram gives valuable information about the shock velocities. A typical trace of shock trajectory from the $x-t$ diagram given in Figure \ref{figure7}b (5th cycle) is extracted digitally through an in-house Matlab program through a combination of the edge-detection algorithm. The three different phases of the pulsation cycle are shown in the extracted trajectory as shown in Figure \ref{figure8}a. One of the applicable parameters of this trajectory is that the gradient of it will lead directly to the velocity of shock motion as shown in Figure \ref{figure8}b. After the analysis of Figure \ref{figure8}, it is evident that the shock system accelerates rapidly against the flow direction up to a $[u/U_\infty]\approx 1$ during part of the `inflation' stage and then decelerates as it nears the spike tip. Later, due to the `with-hold' stage, the shock stays constant but expands in the transverse direction. Due to it, the velocity values are observed to be zero. At the end of the `with-hold' stage, the shock system collapses and compresses the recirculation gas against the forebody. At the time of the `collapse' stage, the shock system accelerates and achieves a maximum velocity of $[u/U_\infty]\approx 0.25$ before the next cycle starts. These analyses stay consistent with the numerical findings of Feszty \etal \citep{Feszty2004a}.

\subsection{Unsteady pressure signals}\label{sec:unsteady_p}

\begin{figure*}
	\centering \includegraphics[width=0.8\textwidth]{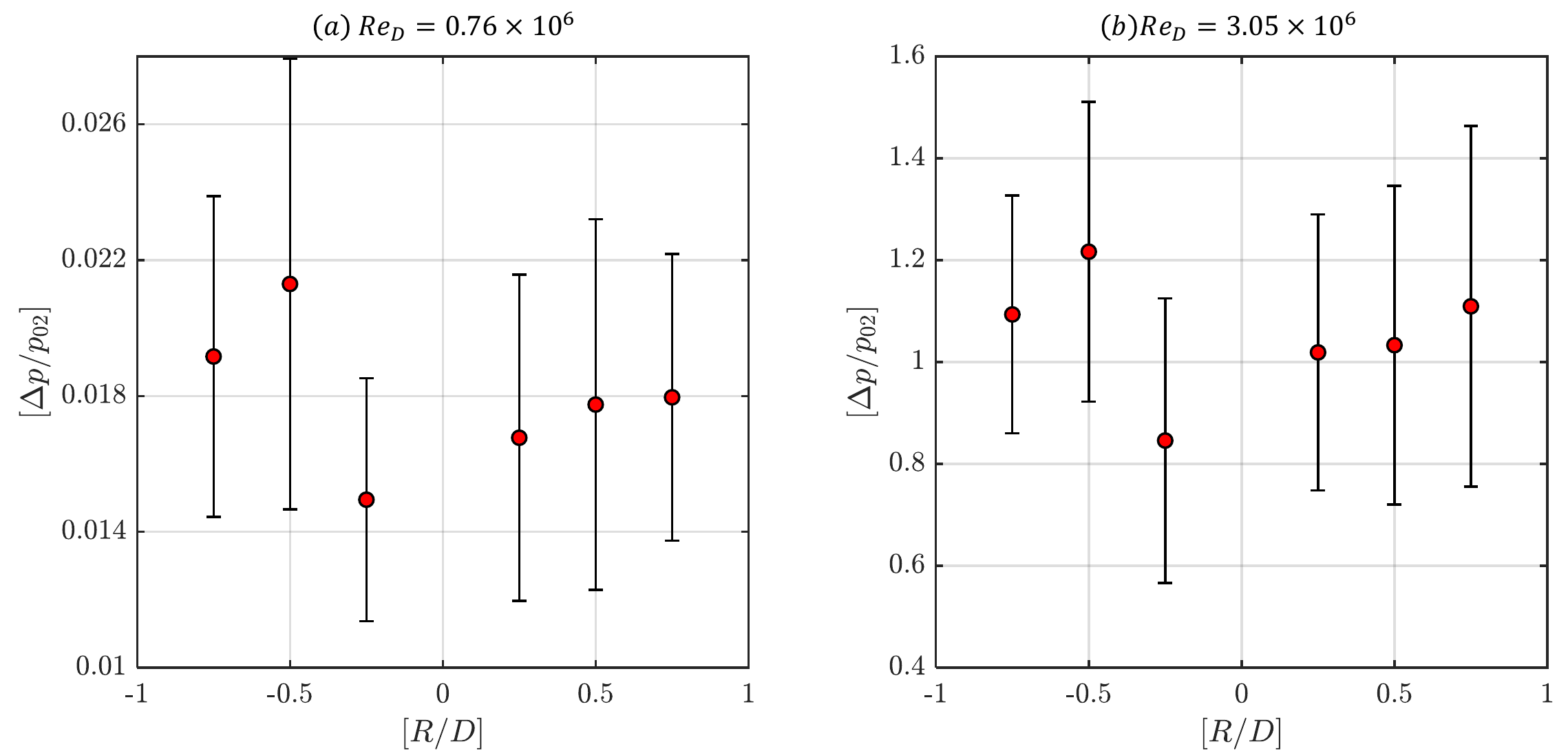}
	\caption{Normalized pressure measurements on the flat-face spiked body flow showing the $rms$ ($\Delta p_{rms}/p_{02}$, red filled circles) and fluctuation intensity ($\sqrt{\overline{\Delta p^2}}/p_{02}$, error-bar) for two different cases of $Re_D$ at the same freestream Mach number of $M_\infty=8.16$: (a). $Re_D = 0.76 \times 10^6$ (low), and (b). $3.01 \times 10^6$ (high).}
	\label{figure9}
\end{figure*}

\sk{In order to understand the underlying flow physics of the pulsation cycle clearly, unsteady pressure signals are acquired at six vital locations (see Figure \ref{figure1}b) symmetric about the axis. A typical pressure cycle observed during pulsation is shown in Figure \ref{figure6}a with all the three stages distinctly marked for high $Re_D$ case. The timestamp of solid red dots represents the time stamp of the instantaneous schlieren snapshots given in Figure \ref{figure7} for one-to-one comparison. In Figure \ref{figure9}, the root-mean-square ($rms$) of the pressure variation (pressure loading) is given as solid red-dots and the length of error-bar represent the pressure fluctuation intensity ($\sqrt{\overline{(p-\overline{p})^2}}=\sqrt{\overline{(\Delta p)^2}}$) for two different cases of $Re_D$. A maximum drop of 98.24\% is seen between the high and low $Re_D$ cases in pressure loading and fluctuation intensity.} From the time-averaged schlieren image given in Figure \ref{figure4}b, the flow looks almost symmetric, and the angle of attack of the model was indeed set to $0^\circ$ with $\pm 0.1^\circ$. However, the pressure distribution in Figure \ref{figure9}b is not symmetric. One of the primary reasons for this behavior is explained from the point of view of radial shock-related instabilities, which produces rotating stationary waves\citep{Demetriades1985}. The phenomena mentioned above is a topic by itself, and it will not be considered for further discussion.

\begin{figure*}
	\centering \includegraphics[width=0.7\textwidth]{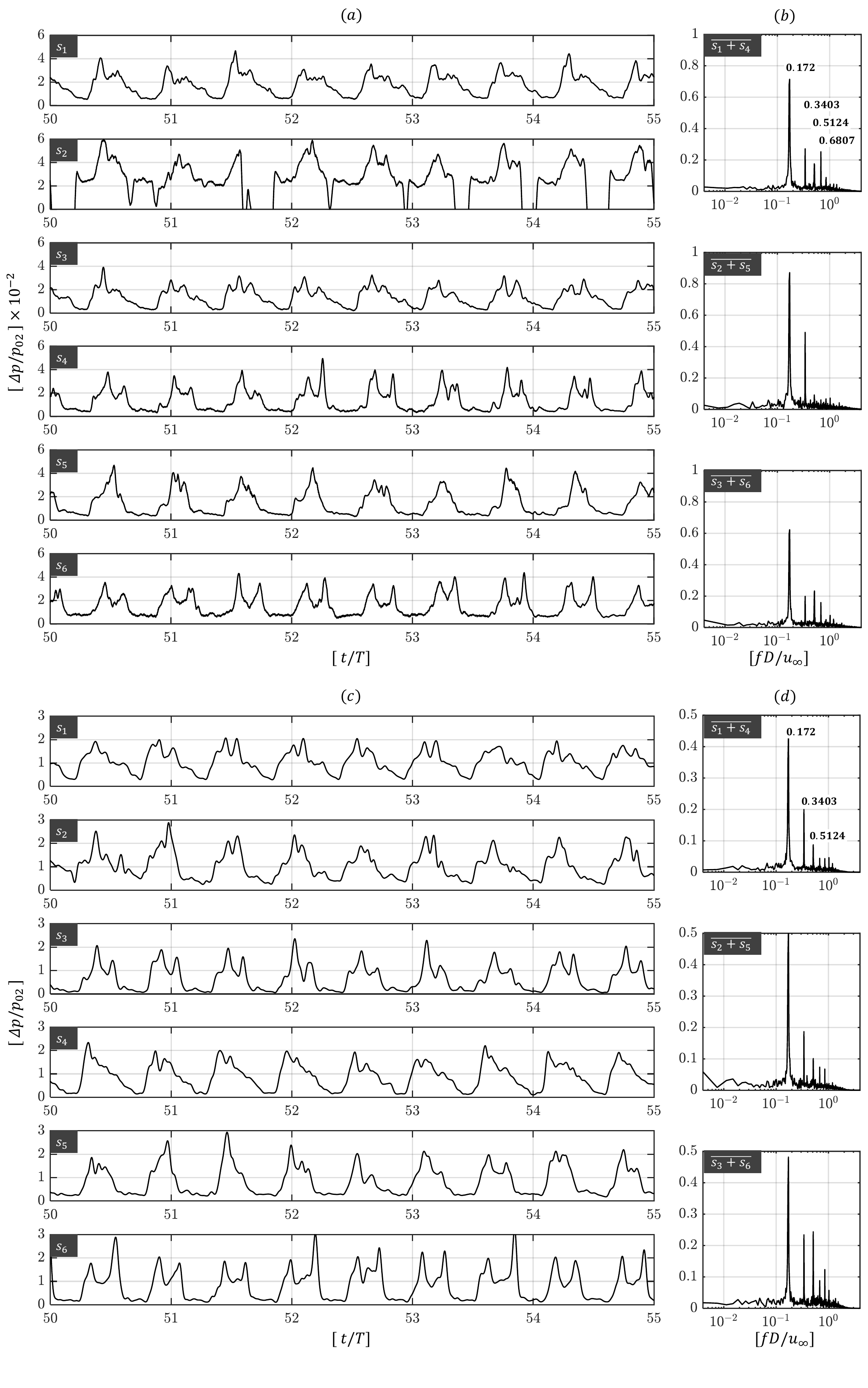}
	\caption{(a, c) Pressure cycles observed at individual locations from $s_1$ to $s_6$ (see Figure \ref{figure1}b) during the run-time for two different $Re_D$ cases; (b, d) \sk{Fast-Fourier Transform (FFT) of the spectrally averaged pressure signals (considering the symmetry of sensor placement) showing the presence of discrete dimensionless frequencies and normalized pressure amplitudes ($\Delta p/p_{02}$)} for two different $Re_D$ cases.}
	\label{figure10}
\end{figure*}

A typical trace of pressure history from 10-cycles for each of the sensors (from $s_1$ to $s_6$) is given in Figure \ref{figure10}a and Figure \ref{figure10}c for both $Re_D$ cases. The asymmetric variations in amplitude and a small phase shift in the signals from the sensors placed symmetrically about the axis can be easily observed due to radial shock-related instabilities as mentioned earlier. \sk{The FFT of the signals from the symmetrically placed sensors after spectral averaging ($\overline{s_1+s_4}, \overline{s_2+s_5}$ and $\overline{s_3+s_6}$) is given in Figure \ref{figure10}b and Figure \ref{figure10}d to study the dominant frequencies involved in the pulsation cycle and also to avoid cluttering of similar figures for different $Re_D$. The fundamental is observed at $[f_1D/u_\infty]=0.172$, and the overtones are observed at $[f_2D/u_\infty]=0.3403$ and $[f_3D/u_\infty]=0.5124$, respectively exactly in the relation of $[f_3D/u_\infty]=3[f_1D/u_\infty]$ and $[f_2D/u_\infty]=2[f_1D/u_\infty]$ as seen in the $x-t$ diagram analysis of Figure \ref{figure7}c.} These similarities will help in corroborating the findings from the schlieren images and unsteady pressure signals. In addition, from the empirical relation given by Kenworthy \cite{Kenworthy1978} as given below,
\begin{equation} \label{eq:kenworthy}
   \left(\frac{f_e D}{u_\infty}\right) = \left[0.25-0.067\left(\frac{l}{D}\right)\right]
\end{equation}
after substituting the values given in Table \ref{table1}, and Figure \ref{figure3}, the estimated frequency, $[f_eD/u_\infty]=0.171$, which is almost closer to $[f_1D/u_\infty]$ from the unsteady pressure measurements by only $\pm0.6$\%.

Unlike the earlier findings by Feszty \etal \citep{Feszty2004a}, where it was observed that the gas trapped inside the vortical region V2, causing the rapid expansion, it is clear that it is not just the vortical region V2 near the root of the spike but also the growing vertical region V1, contributes to the rapid expansion. The interaction of the votrical regions (V1 and V2 and with the wall) followed by the rapid release of the trapped gas inside them (as seen in Figure \ref{figure5} of frame \textcircled{m} at $\tau+12\Delta \tau$) results in pressure rise near the root of the spike as seen in Figure \ref{figure6}, where the pressure peaks to $[\Delta p/p_{02}]\sim 2$. The first peak appears at the interaction of the approaching shock and the shock arising due to the compression of collapsing gas against the forebody. The second peak in the pressure cycle arises from the refraction of approaching shock against the forebody wall, which leads to the continuation of the next cycle. The interaction of shock systems sheds interacted vortices back into the recirculation region, which causes rapid inflation. The inflation decreases pressure on the flat-face, as seen in the pressure cycle. At the end of the inflation stage, the pressure curve plateaus for a specific duration and begins to fall further, representing the with-hold stage.    

\subsection{Modal analysis} \label{sec:modal_analysis}

The dynamic events happening during the pulsation cycle can be better understood through the modal analysis \citep{Kutz2016}. Two modal decomposition tools are employed to understand the shock motion: a. Proper orthogonal decomposition (POD, $\Phi(x/D,y/D,a_\phi(t/T))$), and b. Dynamic mode decomposition (DMD, $\Theta(x/D,y/D,t/T)$). The first method helps understand the energy contents contained in each of the modes and the total number of dominant modes required to represent the flowfield. The second method will help identify the dominant temporal contents and the corresponding spatial modes.

The method involves the preparation of images that will carry only the fluctuations from the fluid phenomena but not the artifacts from the anomalies due to instrumentation. The images are prepared using the procedures mentioned in Karthick \etal \citep{Karthick2017} and all the unwanted features like parasite reflections, window defects, and light spots are removed. The processed images are loaded into a column matrix, and 1000 such images are used to construct the complete matrix. Later, through a series of single value and eigenvalue decomposition, POD and DMD modes are extracted by using the procedure thoroughly explained in the book of Kutz \etal \citep{Kutz2016}. \sk{Due to the limited intensity fluctuations seen in the high-speed schlieren images of low $Re_D$ case, the entire set is discarded for modal analysis as no valuable conclusions could be drawn. On the other hand, the high $Re_D$ case is considered given that there are distinct flow features present in the respective high-speed schlieren images.}

\begin{figure*}
	\centering \includegraphics[width=\textwidth]{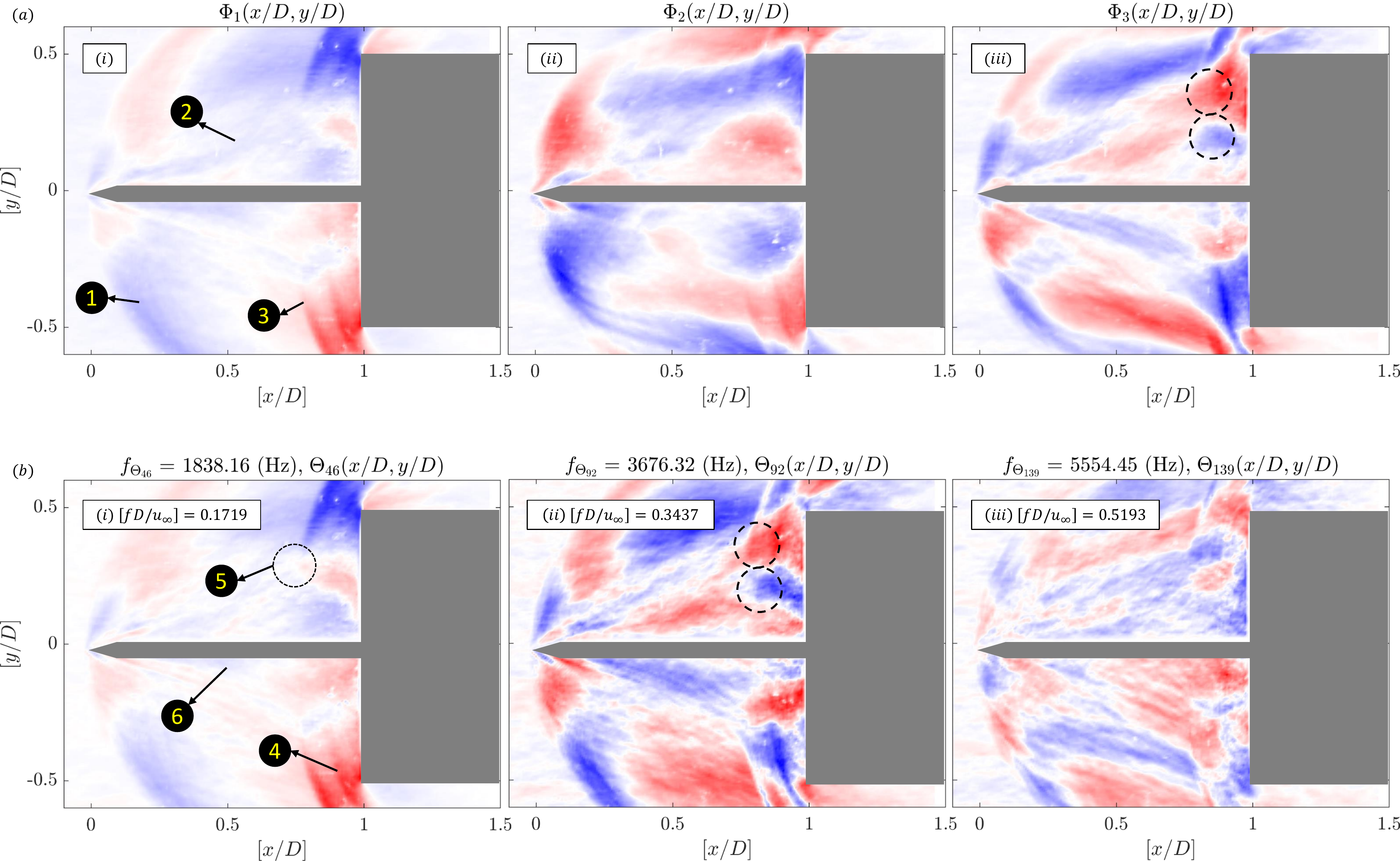}
	\caption{\sk{Contour plots showing the similarity between the (a) first three dominant POD spatial modes ($\Phi_{1-3}(x/D,y/D)$) and (b) the first three dominant DMD spatial modes: (i) $[f_{\Theta_{46}}D/u_\infty]=0.1719$, (ii) $[f_{\Theta_92}D/u_\infty]=0.3437$, and (iii) $[f_{\Theta_{139}}D/u_\infty]=0.5193$. Dominant flow features: 1. inflated shock in the leading edge of the spike; 2. separated free shear layer; 3. Collapsed shock forming ahead of the flat-face; 4. Shock formation due to the rapid compression of collapsing flowfield; 5. Triple point; 6. Re-circulation region.}}
	\label{figure12}
\end{figure*}

\begin{figure*}
	\centering \includegraphics[width=0.7\textwidth]{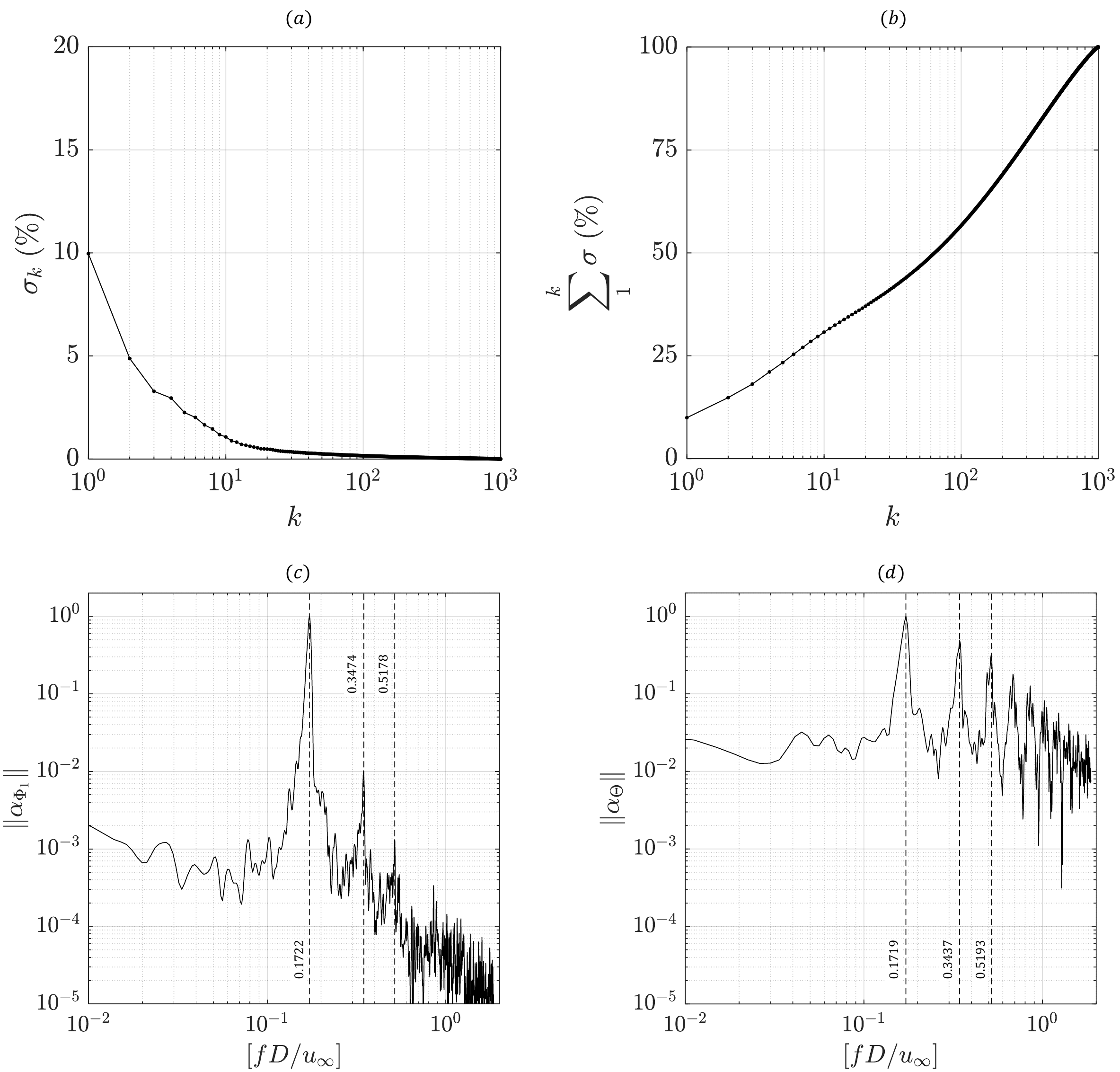}
	\caption{\sk{(a) Plot showing the variation of energy contents ($\sigma_k$) in each of the POD ($\Phi(x/D,y/D)$) modes ($k$); (b) Plot showing the cumulative energy distribution from the POD ($\Phi(x/D,y/D)$) analysis. (c) Plot showing the presence of discrete frequencies ($[f_{\Phi}D/u_\infty]$) at different normalized amplitude ($||a_{\Phi_1}||$) from the temporal coefficients variation of the dominant POD temporal mode ($a(\Phi_1(t)$); (d) Plot showing the presence of discrete frequencies ($[f_{\Theta}D/u_\infty]$) at different normalized amplitude ($||\alpha(\Theta)||$) from the DMD temporal mode ($\Theta(t/T)$) analysis.}}
	\label{figure13}
\end{figure*}

In Figure \ref{figure12}a and Figure \ref{figure13}c, the dominant energetic spatial mode ($\Phi_1(x/D,y/D)$) and the \sk{FFT} of the dominant energetic temporal mode is given. Looking at the spatial mode (Figure \ref{figure12}a), it can be seen that the features represent the time-averaged flow features as seen in Figure \ref{figure4}b of the schlieren image. Thus, it can be concluded that the shock motion in the extremities of the spike tip forms the dominant flow features. The FFT from the temporal mode (Figure \ref{figure13}c) also in agreement with $x-t$ diagram's \sk{FFT} (Figure \ref{figure8}c) and the \sk{FFT} of unsteady pressure signal (Figure \ref{figure10}d).

From the POD analysis, the energy contents in individual modes and the cumulative energy distribution are given in Figure \ref{figure13}a-b. It can be seen that it requires only the first six modes to represent the 25\% of flow energy (derived from the fluctuation square of the density gradients). \sk{Among them, the first mode ($\Phi_1(x/D,y/D)$) alone contains about 10\% of the total energy, whereas the second ($\Phi_2(x/D,y/D)$) and third ($\Phi_3(x/D,y/D)$) spatial modes represents 5\% and 3\% of the total energy, respectively. From analyzing the $\Phi_2(x/D,y/D)$ and $\Phi_3(x/D,y/D)$, the structures represent the inflation phase and the merging of vortices during the collapse phase. The interpretation is drawn by comparing the most distinct spatial features from the energetic spatial modes with that of the instantaneous schlieren images as shown in Figure \ref{figure5}.}  

From the DMD analysis, the dominant dynamic spatial mode (\ref{figure12}b) provides information on the spatial extent of shock oscillation. The corresponding oscillation frequency is given in the dominant DMD temporal mode \ref{figure13}d. These findings shed similar information as seen from the $x-t$ diagram and the unsteady pressure analysis as before. Since the dominant dynamics are from the shock-motion, the dynamic spatial mode ($\Theta_{46}(x/D,y/D)$) corresponding to $[f_{46}D/u_\infty]=0.1719$ as shown in Figure \ref{figure12}b-(i) represents the same time-averaged flow feature in Figure \ref{figure4}b and the POD dominant energetic spatial mode (see Figure \ref{figure12}a-(i)). However, due to the temporal fluctuations, some noise will be observed in the spatial mode. However, for the second dominant temporal mode at $[f_{92}D/u_\infty]=0.3437$, the formation of distinct vortical regions can be seen in Figure \ref{figure13}b-(ii) (shown in dotted circles). Similarly, for the third dominant temporal mode $[f_{139}D/u_\infty]=0.5193$, the corresponding spatial mode in Figure \ref{figure13}b-(iii) shows some parts from each division of the cycle, and it is unclear due to large noise.

From the modal analysis, it is clear that the interaction of the V1 and V2 vortical zone drives the shock oscillation. The burst of these structures is observed at a frequency precisely equal to twice the fundamental frequency of the pulsation.   

\section{Conclusions} \label{sec:conclusion}
\sk{An experimental campaign is carried out to study the pulsating flowfield observed around an axisymmetric flat-face cylinder at zero angles of attack. The experiments are performed in a hypersonic flow generated using the recently converted short-duration hypersonic shock tunnel into a Ludwig tunnel operating for a longer duration. Experiments are done at two different freestream $Re_D$ (0.76 and 3.01 $\times 10^6$) at a constant $M_\infty$. High-speed schlieren imaging, unsteady pressure measurements, $x-t$ analysis, spectral analysis, and modal analysis are performed to understand the flow physics. Following are the major conclusions of the present study:
\begin{enumerate}
	\item From the analysis of high-speed schlieren images at high $Re_D$, the presence of V1 and V2 vortical zones are identified to drive the collapse stage of the pulsation cycle, which is not observed before experimentally.
	\item From the $x-t$ diagram, the frequencies of the vortices interactions are identified, and the presence of V1 and V2 vortices are found to be particularly unique to flows with high $M_\infty$, primarily due to the presence of shallow shock angle and oblique shock impingement on the base body.
	\item From the unsteady pressure signal, a maximum drop of 98.24\% in pressure loading and fluctuation intensity is seen between the different $Re_D$ cases. Furthermore, the dominant pressure load exists at the time of vortex burst and forebody shock formation.
	\item The spectral characteristics observed between the two different $Re_D$ cases reveal invariant dominant frequencies, however, with varying amplitudes and different decay rates. The low and high $Re_D$ cases exhibit an inverse and -7/3 decay rate, respectively, at higher frequencies. The observation indicates the formation of possible turbulent structures at high $Re_D$ during vortical breakdowns.
	\item  From the modal analysis, the third dominant energetic mode and the second dominant dynamic mode comprise the vortical interactions of V1 and V2. The interactions are observed from the DMD analysis at a dimensionless frequency of exactly twice that of the dominant pulsation frequency.
\end{enumerate}}

\sk{The findings will help prepare the forebody shielding with appropriate material to avoid acoustic loads or design an efficient active or passive control device to overcome the unsteadiness. The outcome will also be helpful to come up with new geometrical changes to spike or fore-body shape that can modify or diminish these dominant flow patterns and reduce unsteadiness.}

\section*{Supplemental material}
\sk{The combined vorticity and density contours in the form of a high-speed \href{https://youtu.be/dEvM51VeDCw}{`video'} is given in the supplementary, and it shows the shedding of vortical structures from the $\lambda$-shock's triple point in a pulsating spiked body at hypersonic flow.}

\section*{Authors' Contributions}
M.I.S., R.S., S.K.K., and G.J. have contributed equally to this work. The authors report no conflicts of interest.

\section*{Acknowledgement}
The authors thank the assistance offered by the late assistant Mr. Jeevan of LHSR, IISc-Bengaluru in preparing and instrumenting the models for testing.    

\section*{Data availability statement}
The data that support the findings of this study are available from the corresponding author upon reasonable request.

\section*{References}

\bibliography{references}
\onecolumngrid
\PRLsep	
\end{document}